\documentclass[runningheads]{llncs}
\usepackage[export]{adjustbox}
\usepackage{graphicx}
\usepackage{url}
\usepackage{float}
 \usepackage{listings, paralist}
 \PassOptionsToPackage{hyphens}{url}
 \usepackage{hyperref}

\usepackage[T1]{fontenc}
\usepackage[scaled=0.85]{beramono}
\usepackage{listings}
\usepackage{comment}
\usepackage{todonotes}
\usepackage{wrapfig}

\begin{document}
\title{Making Study Populations Visible through Knowledge Graphs}
\author{Shruthi Chari \inst{1}\orcidID{0000-0003-2946-7870} \and
Miao Qi \inst{1}\orcidID{0000-0002-2917-0965} \and Nkechinyere N. Agu \inst{1} \orcidID{0000-0003-1386-8602} \and
Oshani Seneviratne \inst{1}\orcidID{0000-0001-8518-917X} \and James P. McCusker \inst{1}\orcidID{0000-0003-1085-6059} \and   Kristin P. Bennett \inst{1}\orcidID{0000-0002-8782-105X} \and Amar K. Das \inst{2}\orcidID{0000-0003-3556-0844}  \and Deborah L. McGuinness \inst{1}\orcidID{0000-0001-7037-4567} }

\authorrunning{S. Chari et al.}

\institute{Rensselaer Polytechnic Institute, Troy, NY 12180, USA \and
IBM Research, Cambridge, MA
\email{\{charis,qim,agun,senevo,mccusj2,bennek\}@rpi.edu}, amardas@us.ibm.com, dlm@cs.rpi.edu}

\maketitle              

\vspace{-1.5em}
\begin{abstract}
Treatment recommendations within Clinical Practice Guidelines (CPGs) are largely based on findings from clinical trials and case studies, referred to here as research studies, that are often based on highly selective clinical populations, referred to here as study cohorts. When medical practitioners apply CPG recommendations, they need to understand how well their patient population matches the characteristics of those in the study cohort, and thus are confronted with the challenges of locating the study cohort information and making an analytic comparison. To address these challenges, we develop an ontology-enabled prototype system, which exposes the population descriptions in research studies in a declarative manner, with the ultimate goal of allowing medical practitioners to better understand the applicability and generalizability of treatment recommendations. We build a Study Cohort Ontology (SCO) to encode the vocabulary of study population descriptions, that are often reported in the first table in the published work, thus they are often referred to as Table 1. We leverage the well-used Semanticscience Integrated Ontology (SIO) for defining property associations between classes. Further, we model the key components of Table 1s, i.e., collections of study subjects, subject characteristics, and statistical measures in RDF knowledge graphs. We design scenarios for medical practitioners to perform population analysis, and generate cohort similarity visualizations to determine the applicability of a study population to the clinical population of interest. Our semantic approach to make study populations visible, by standardized representations of Table 1s, allows users to quickly derive clinically relevant inferences about study populations.\\

\textbf{Resource Website:} \url{https://tetherless-world.github.io/study-cohort-ontology/}

\end{abstract}

\section{Introduction} \label{introduction}
\begin{figure}[hbt!]
  \includegraphics[width=1.0\linewidth]{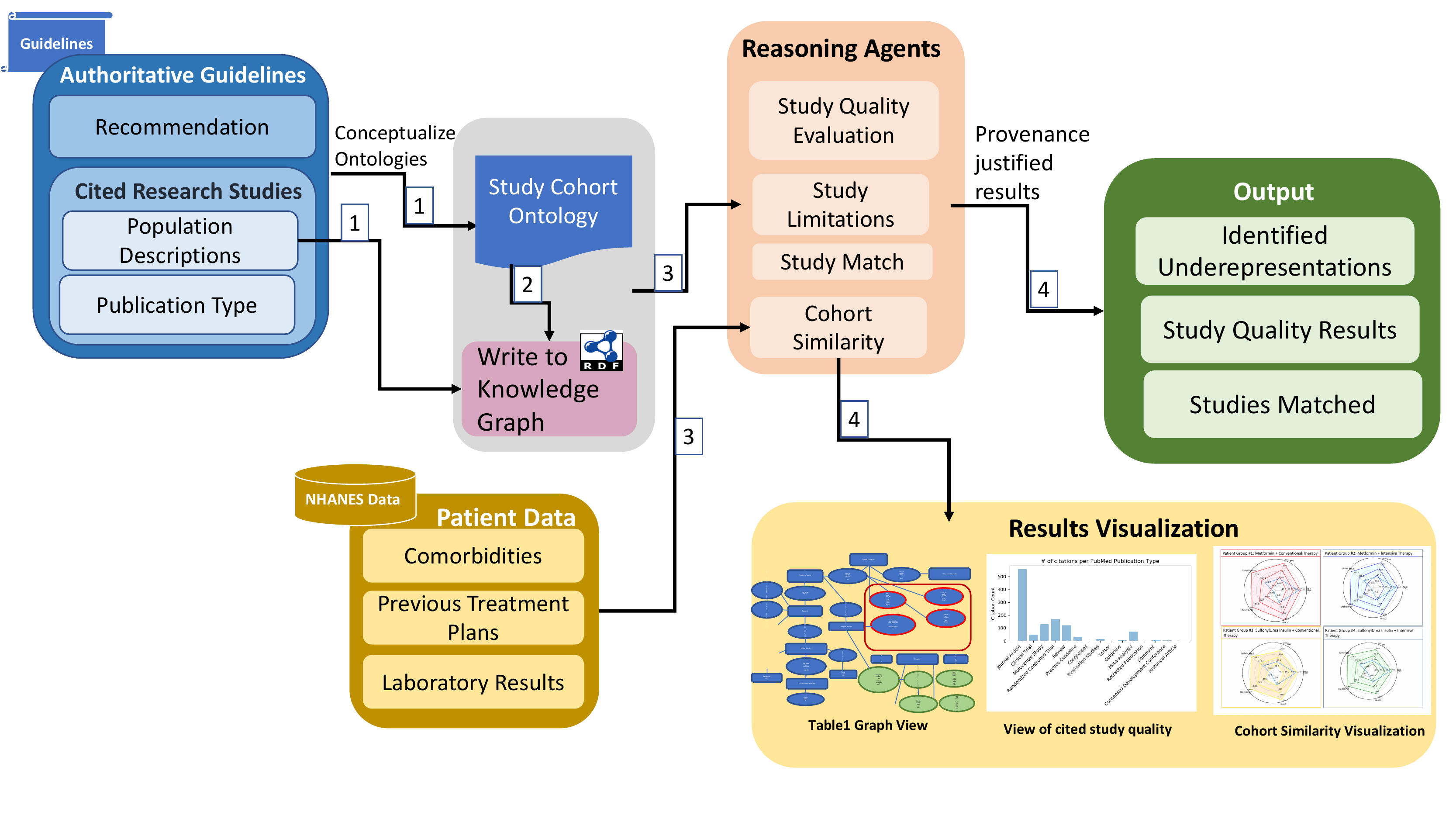}
  \caption{An overview of the cohort analytics workflow which 1) ingests terms from population descriptions of research studies, 2) standardizes their representations via KR techniques and 3) supports study applicability applications. The numbering is in-line with the figure and is indicative of data flow.}
  \label{cohort analytics workflow}
\end{figure}

Our goal is to build a semantic solution to model the descriptions of study populations and to assist medical practitioners in determining the applicability of a study to their clinical population. Through Fig. \ref{cohort analytics workflow}, we describe the components of a prototype system, that utilizes knowledge representation (KR) techniques to model tabular representations of study population descriptions, often captured in the first table of the scientific publication. We build a Study Cohort Ontology (SCO) (section \ref{ontology modeling}) to support the vocabulary in these Table 1s (plural form) and to model their structure. Further, we encode Table 1s as Resource Description Framework (RDF) knowledge graphs (KGs)~\cite{auer2018towards} (section \ref{kg}) to expose in a declarative manner\footnote{declarative manner: in a clear, unambiguous, and computer understandable manner} these study populations. We demonstrate our ontology and the use of our knowledge graphs with two applications (section \ref{applications}): one aimed at helping medical practitioners determine the similarity of a patient or a clinical population to the study population, and another aimed at supporting retrospective analysis of a study to expose possible biases or population gaps, such as racial underrepresentations.   

\subsection{Use Case} \label{motivation}
Evidence-based Medicine (EBM) has been gaining popularity, and medical practitioners are using it more often. However, it is challenging to design the CPGs to stay current with the growing body of clinical literature. Additionally, medical literature is continuously being revised, e.g., typically, new versions of CPGs are released annually. Treatment recommendations in CPGs are often supported by evidence from cited research studies, i.e. clinical trials and observational case studies, targeting highly selective populations with sociodemographic and comorbid characteristics. In clinical practice, it is well-known that there are biases in clinical evidence that reduce their generalizability. The widely-cited research article, ``Trustworthy Clinical Practice Guidelines: Challenges and Potential,'' \cite{graham2011trustworthy} states some of the problems in existing guideline practices, such as ``Failure to include major population subgroups in the evidence base thwarts our ability to develop clinically relevant, valid guidelines.''\\
Furthermore, when medical practitioners are faced with the treatment of complicated patients who do not wholly align with guideline recommendations, they 
may want to consult research studies with relevant findings to determine if the study applies to their clinical population. Hence, we are developing a semantic solution to address these challenges, by providing medical practitioners access to high-quality and applicable guideline evidence. We evaluate our solution on the American Diabetes Association's (ADA) Standards of Medical Care 2018 CPG \footnote{ADA 2018 CPG at: \url{https://diabetesed.net/wp-content/uploads/2017/12/2018-ADA-Standards-of-Care.pdf}} cited research studies, which we will introduce in section \ref{dataset}. 
\section{Related Work} \label{related work}
Existing ontologies for study design and clinical trials are more focused on the study design and methodology aspects of clinical trials, and their vocabulary is insufficient to support cohort descriptions. ProvCaRe \cite{provcare}, an ``Ontology for provenance + healthcare research,'' was developed to assess the scientific rigor and reproducibility of scientific literature. Based on the NIH ``Rigor and Reproducibility'' guidelines \cite{nihrigorandreproducibility},this ontology identifies three components of a study contributing to provenance: study methods (study protocol followed), study instruments (equipment and software used in the study),and study data (metadata about data collection). However, within the ProvCaRe ontology, support for study data is limited to that of inclusion and exclusion criteria, and there is no support for Table 1 terminology, such as subject characteristics and study arms. The Ontology of Clinical Research (OCRE) \cite{sim2014ontology}, a widely cited study design ontology used to model the study lifecycle, addresses goals similar to our study applicability scenario. They adopt an Eligibility Rule Grammar and Ontology (ERGO) \cite{tu2011practical} annotation approach for modeling study eligibility criteria to enable matching a study's phenotype against patient data.\\ 
Since we encode a provenance component of guideline evidence, we searched for 
ontologies for scientific publications. We found that most clinical trial ontologies, e.g., CTO-NDD \cite{younesi2014knowledge}, are domain specific and not directly reusable for a population modeling scenario. Other ontologies, such as the EPOCH
suite of clinical trial ontologies \cite{shankar2006epoch}, that was developed to track patients through their clinical trial visits, had class hierarchies that were insufficient to represent the types of publications cited in the ADA Standards of Care CPG. Additionally, there is another cohort ontology \cite{thecohortontology} being developed. However, our modeling of the association of descriptive statistics with subject characteristics differs from their modeling decision to define new properties to represent these associations. Instead, we introduce classes to accommodate new subject characteristic terms upon Table 1 ingestion, and we limit the number of descriptive statistics to a standard set of central tendency measures and boundary values. Hence, we do not leverage their ontology. Further, their ontology is domain specific, including many sleep disorder classes. In SCO we provide a generalized and richer, domain-agnostic Table 1 vocabulary (sufficient to support research studies targeting various diseases).\\
Clinical trial matching has been attempted multiple times, largely as a Natural Language Processing problem, including a KR approach that improves the quality of the cohort selection process for clinical trials \cite{patel2007matching}. Clinical trial matching work \cite{patel2007matching} was carried out with the help of an ontology, and TBOX (knowledge-based) assertions were created from SNOMED-CT for supporting ABOX (real-world) assertions of patient records. However, the focus of their effort was mainly on efficient KR of patient data, and study eligibility criteria was formulated as SPARQL queries on the patient schema. We tackle the converse problem of identifying studies that are applicable to a clinical population based on the study populations reported. We address this problem from the perspective of modeling the study populations. 
\vspace{-0.4cm}
\section{Dataset} \label{dataset}
\vspace{-0.1cm}
Our evaluation dataset is comprised of research studies, cited in the ADA Standards of Medical Care 2018 CPG. 
We manually reviewed the entire guideline to understand the types of evidence utilized to support treatment recommendations. 
ADA treatment recommendations are supported through citations within the discussion, which serve as implicit evidence for the recommendation.
Further, we used PubMed APIs\footnote{\url{https://pypi.org/project/pubmed-lookup/}}
on the Medline\footnote{\url{https://www.nlm.nih.gov/bsd/medline.html}} publications, cited in evidence sentences across chapters of the ADA CPG, to retain only those publications that met the qualifications for our definition of research studies. We only considered publications tagged with Pubmed Publication types\footnote{Find the list of all supported publication types at \url{https://www.ncbi.nlm.nih.gov/books/NBK3827/table/pubmedhelp.T.publication_types/}} of: Randomized Controlled Trial, Clinical Trial, and Multicenter Study.\\
We focused on the pharmaceutical treatments and comorbidities associated with type-2 diabetes, and we  filtered our evaluation dataset to contain cited research studies from the Pharmacologic Interventions (Chapter 8) \cite{ada20188} and the Cardiovascular Complications (Chapter 9) \cite{ada20189} of the ADA 2018 CPG.
We did a thorough, manual investigation of research studies from these chapters, looking for any variance in Table 1s and identifying important study data that explained Table 1 variables. Furthermore, although we were able to gather full-text links for Medline citations through programmatic means, we had to manually follow these links to ensure they are freely available, and, if not, we checked for the availability of the study in other sources. Due to these challenges, we narrowed down the number of research studies to 20 that we list on our resources website.
\section{Study Cohort Ontology} \label{ontology modeling}
As introduced in section \ref{introduction}, we build a Study Cohort Ontology (SCO) to serve as a vocabulary to model the components of a Table 1, the study arms (columns) and their characteristics (rows). We also ensure that the implicit associations exhibited between these components are reflected in SCO. We adopt a bottom-up approach to modeling, that follows, as a by-product of our investigative efforts, the description in section \ref{dataset}. Further, we have attempted to keep our main SCO ontology as domain-agnostic as possible to ensure easy reuse and longevity. In subsection \ref{main classes and properties}, we introduce the main concepts in our ontology to provide a contextual understanding of 
the descriptions of populations reported in Table 1s, and walk through our approach to ontology reuse in subsection \ref{ontologies reused}.

\begin{figure}[hbt!]
  \includegraphics[width=1.10\linewidth,center]{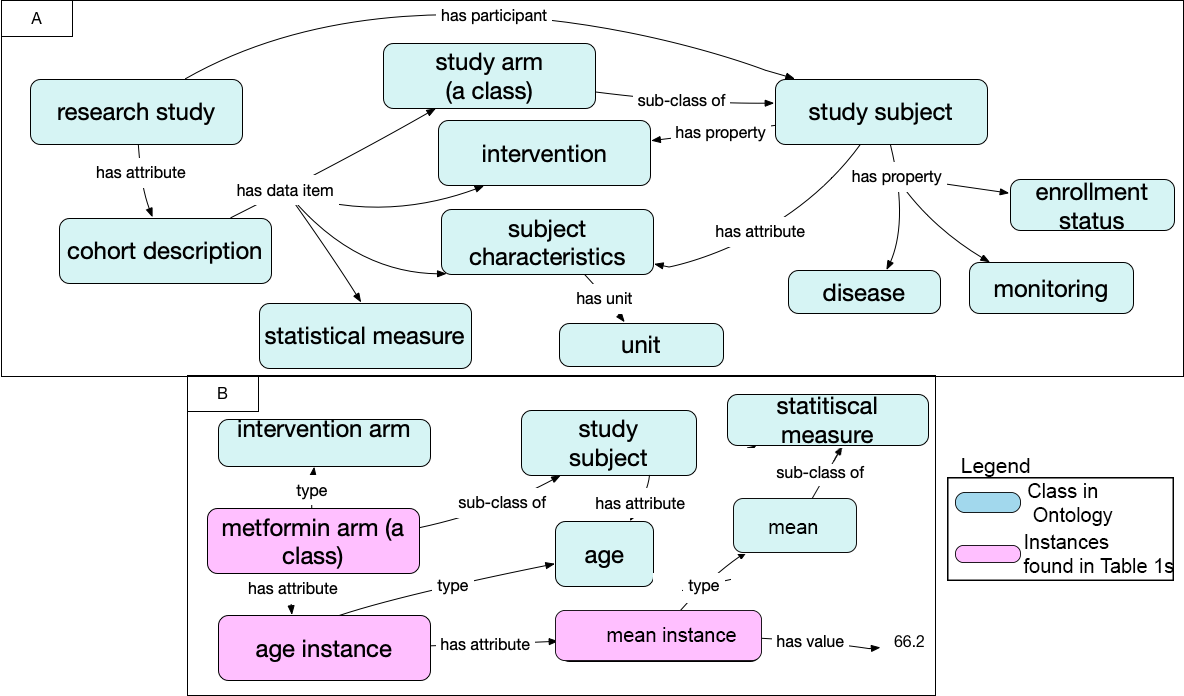}
  \caption{A) A high-level overview of SCO that captures the vocabulary and associations needed to model the descriptions of study populations. B) We depict associations that cannot be realized without actual instantiation of Table 1 data.}
  \label{concept map}
    \end{figure}
\subsection{
Primary Classes and Property Associations} \label{main classes and properties}
The descriptions of study populations that are reported in Table 1s follow a pattern in which columns represent study arms, a group of study subjects 
who receive an intervention or control regime. The subject characteristics are presented in rows, and are aggregated upon and reported via descriptive statistical measures in the cells of the table. In a conceptual model of SCO as shown in Fig. \ref{concept map}, we depict our modeling of these Table
1 components and the additional details that are necessary to describe a study population in the context of a research study. A more detailed version can be found on our resources website.\\
As will become evident from a representative Table 1 example shown in Fig. \ref{table 1 example 2}, the row and column headers in Table 1s
contain specific medical codes and variables that can further be grouped into broad general classifications: Anthropometric Properties (chear:Anthropometry),\footnote{
We use the ontology prefixes: (1) sio: SemanticScience Integrated Ontology (2) uo: The Units of Measurement Ontology (3) chear: Children's Health Exposure Analysis Resource Ontology (4) ncit: National Cancer Institute Thesaurus (5) provcare: ProveCaRe (6) doid: Human Disease Ontology (7) sco: Study Cohort Ontology (8) hasco: Human-Aware Science Ontology (9) prov: The PROV ontology (10) dct: Dublin Core Terms (11) vann: A vocabulary for annotating vocabulary descriptions} Demographics (chear:Demographic), Laboratory Results (ncit:C36292), Diseases (doid:0004), and Medical Interventions (provcare:Intervention). 
Further, we associate all these broad, general classifications we just identified, such as subject characteristics, diseases, interventions etc., via sio:hasAttribute and sio:hasProperty relations to the study subject. More specifically, for properties such as disease and interventions that persist over time and are characterized by the state the study subject 
exhibits,\footnote{View the definition of sio:hasProperty and sio:hasAttribute relations at: \url{https://raw.githubusercontent.com/micheldumontier/semanticscience/master/ontology/sio/release/sio-subset-labels.owl}} we use a sub-property of sio:hasAttribute, i.e. sio:hasProperty, to link them to the study subject. 
Additionally, we do not maintain certain property associations (e.g. compositional relation between the study arm and study subject) in our ontology and only create them upon the representation of actual Table 1 content in RDF KGs. For the ease of understanding, we depict instances in pink in Fig. \ref{concept map} to help visualize the realism in our modeling.\\ 
To summarize, essentially through SCO, we build a framework to model a set of study subjects, who participate (sio:isParticipantIn) in a research study and belong to a study arm and whose subject characteristics are measured (sio:hasUnit) in units, and are aggregated upon via descriptive statistics. 
Since we are dealing with the biomedical domain, where multiple definitions may exist for a term, through blank nodes and reification techniques we allow support for this and we maintain provenance for our definitions via prov:wasAttributedTo (person) and dct:source (online source). For example; hasco:ResearchStudy sio:hasAttribute [ a skos:definition; sio:hasValue 'A scientific investigation that involves testing a hypothesis'; prov:wasAttributedTo AmarDas]. Additionally, we also provide example usages of our terms via vann:example, to help future users/contributors of our ontology get an idea of the intended usage of the class. Our main SCO ontology, and our accompanying suite of ontologies, Lab Results, Diseases, Drugs, and Therapies, in which we maintain diabetes specific content, are available as resources. Further, we tested our ontology with the Hermit reasoner.
\subsection{Ontology Reuse} \label{ontologies reused}
We reuse classes and properties from existing biomedical ontologies as much as possible, and only define them ourselves when they do not exist.
We primarily reused ontologies available from Bioportal \cite{noy2009bioportal} that are regularly maintained and have significant reuse.
We have tried to reuse terms from a small set of applicable ontologies to avoid enlarging the ontology when we bring in new classes and additional axioms. We categorize the ontologies, from which we reuse terms, broadly into Study Design ontologies (ProvCaRe, HASCO), Mid-Level ontologies (SIO), Medical ontologies (NCIT, CHEAR, etc.), and Statistical ontologies (STATO, UO). We present a list of our reused ontologies against their groupings on our resources website.\\
In our approach to ontology reuse, we include minimum information to reference  a term (MIREOT) \cite{courtot2011mireot} for most of our reused ontologies, such as ProvCare and NCIT, unless we leverage their structure completely. However, we do import a light-weight version of the Child Health Exposure Analysis and Resource (CHEAR) ontology, by applying the MIREOT technique to extract the demographics and anthropometric branches alone.
We prefer to import the CHEAR ontology, as it builds off SIO and additionally imports 
the HAScO human aware science ontology, that we leverage.
We utilized an online tool, Ontofox \cite{xiang2010ontofox}, to apply the MIREOT technique to a few ontologies that were supported on this platform. However, for ontologies that were not available on Ontofox, we designed our own SPARQL query to gather subclass and superclass trees for a given ontology class. On our resources website, we make  our MIREOT Python script available. This runs the SPARQL query  against a Blazegraph endpoint and returns the RDF version of the subset class tree.

\section{Knowledge Graph Modeling} \label{kg}
\begin{figure}[hbt!]
  \includegraphics[width=0.95\linewidth]{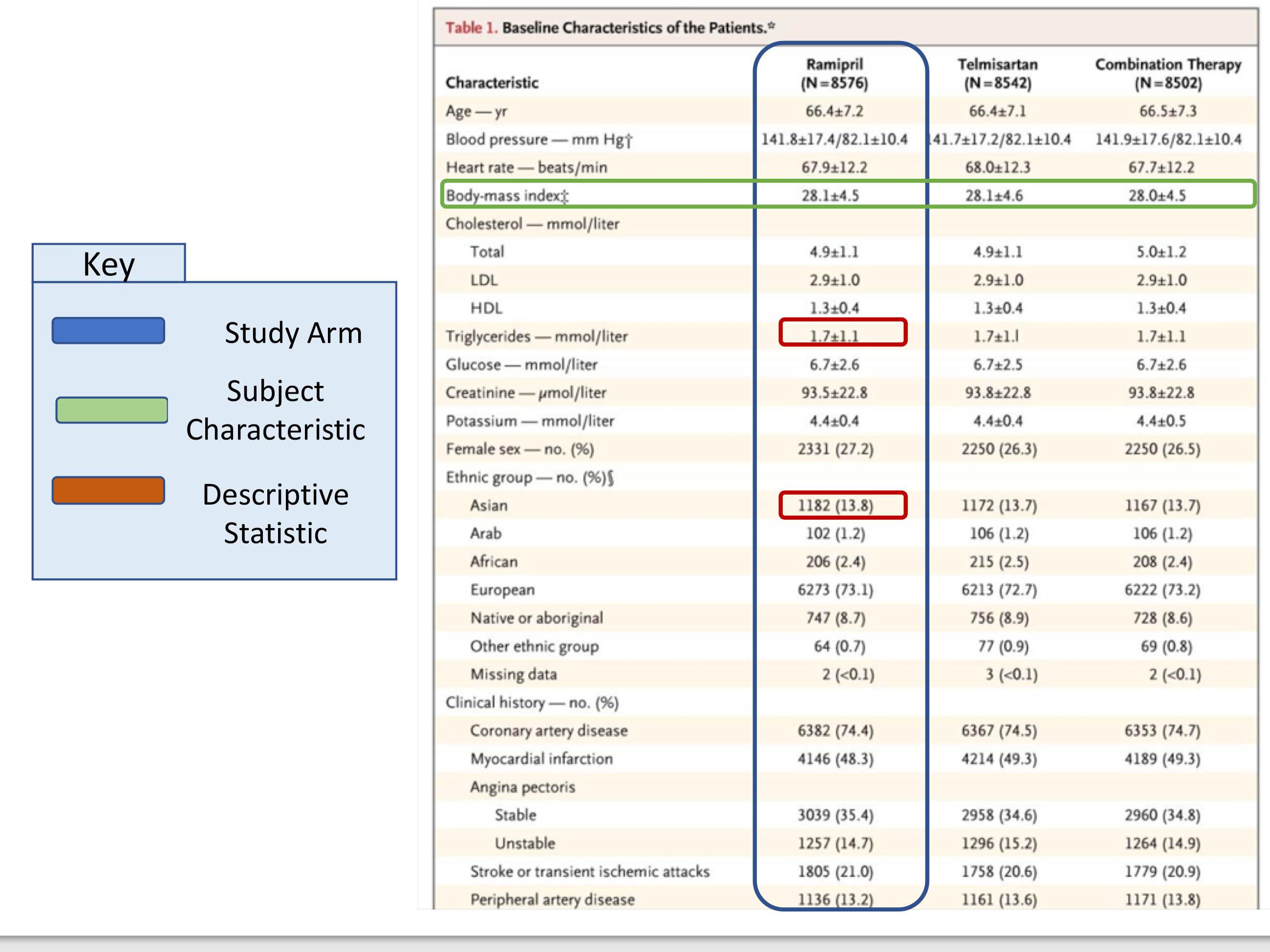}
  \caption{An annotated example of Table 1 from a clinical trial
  ``Telmisartan, ramipril, 
    or both in patients at high risk for vascular events'' \cite{ontarget2008telmisartan} cited in the Cardiovascular Complications (Chapter 9) of the ADA CPG}.
  \label{table 1 example 2}
    \end{figure}

We use an annotated example of a Table 1, seen in Fig. \ref{table 1 example 2}, to explain our approach of modeling the collections of study subjects, subject characteristics defined on collections, and the descriptive statistics used to summarize these characteristics. We present an RDF snippet in Listing 1.1, and explain smaller sub-portions of our modeling in each subsequent subsection. These snippets form the fundamental pieces of our Table 1 KG. On our resources website, we release the KG representations of the studies in our evaluation dataset, for interested readers to run their analyses.
    
\subsection{Modeling of Collections of Study Subjects} \label{collections}

\begin{lstlisting}[frame=tb, caption=Representation of a portion of the Ramipril Study Arm, label=characteristics_example_1,
   basicstyle=\ttfamily]
sco-i:RamiprilArm
       a    owl:Class, sco:InterventionArm; 
       rdfs:subClassOf sio:StudySubject;
       sio:isParticipantIn sco-i:TelmisartanRamiprilStudy;   
       sio:hasAttribute    
       [ a sco:PopulationSize; sio:hasValue 8576],
       [ a sio:Age; sio:hasUnit sio:Year;
          sio:hasAttribute
          [ a sio:Mean; sio:hasValue 66.4],
          [a sio:StandardDeviation; sio:hasValue 7.2 ]  
        ] . 
\end{lstlisting}
Study arms are specific subpopulations of study cohorts comprised of a subset of enrolled study subjects. Hence, they are a natural fit for modeling as classes in the OWL web ontology language \cite{bechhofer2004owl}, ``Classes provide an abstraction mechanism for grouping resources with similar characteristics. Like RDF classes, every OWL class is associated with a set of individuals, called the class extension.'', and model collections as classes.\\
As discussed earlier in subsection \ref{main classes and properties}, study arms are represented as columns in Table 1s. 
Further, the RDF snippet in Listing~\ref{characteristics_example_1} shows a semantic definition of a particular study arm as an instance of the sco:InterventionArm. 
Study arm definitions
are either those of \emph{InterventionArm} or \emph{ControlArm} and they are gathered from the Table 1 columns themselves, if sufficient, if not we consult the study data to find relevant content that describes the arms.\\
In some Table 1s, there also exist subsets of study arms, created by the presence of categorical row variables (e.g. percentage of Asians), expressed in percentages\footnote{More Table 1 reporting style and composition details
at \url{https://prsinfo.clinicaltrials.gov/webinars/module6/resources/BaselineCharacteristics_Handouts.pdf}}. Such subsets are expressed as rdfs:subClassOf the main study arm, and have an owl:Restriction defined on them for membership. An example of the representations of these subsets, can be viewed as a part of the KG creation documentation on our resources website.
\subsection{Modeling of Characteristics and Descriptive Statistics} \label{characteristics}
As briefly introduced in subsection \ref{main classes and properties}, subject characteristics are the phenotype properties that are collected for study subjects. In our evaluation dataset, we have observed that all study arms belonging to a study share the same set of characteristics. However, the range of values for these characteristics differ across study arms depending on their composition. Borrowing from our grouping of characteristics from section \ref{ontology modeling}, we reemphasize that characteristics persisting over a period of time are modeled as sio:hasProperty, and the rest are modeled via sio:hasAttribute property. From this discussion it becomes apparent that our modeling of characteristics on study arms is fairly straightforward and we only utilize two SIO property associations.
In Listing 1.1, we depict the association of age as a sio:hasAttribute of the \emph{Ramipril study arm}. 
Further, characteristics can also be classified broadly as categorical, discreet, and continuous. Categorical characteristics are represented in subsets, and their representation is discussed in the previous subsection.
\subsection{Modeling of Descriptive Statistics}
Another 
problem we address in this paper is the KR of aggregate statistics on subject characteristics of study populations. Although aggregate statistics are reported in multiple domains, 
there has been little work on a convention for supporting the modeling of aggregations in RDF. The support for aggregations in Linked Data is presented in \cite{cyganiak2010semantic}. However, their process is more focused on the publishing of statistical data and the metadata than on the representation of statistical data itself.\\
Descriptive statistics have conventionally been defined, as statistical measures that summarize the data.\footnote{Definition adapted from: \url{https://en.wikipedia.org/wiki/Descriptive_statistics}} In Table 1s, they are used to describe summarized values of the characteristics of study subjects, who belong to a study arm. From our analysis of Table 1s, we have seen a limited set of descriptive statistics measures: mean +/- standard deviation, median +/- interquartile range, and percentages.
We model these aggregations and descriptive statistics, seen as reified triples on a property. Reification is an RDF technique developed to ``make statements about statements'' \cite{rdfreification}. 
As can be seen in the RDF snippet above, we define descriptive statistics as reified triples on an age characteristic.
Additionally, since we only reuse SIO object and data properties, we eliminate the need for further punning techniques, to represent these descriptive statistic properties as instances of sio:hasAttribute. In this paper, we only present an example of a mean +/- standard deviation measure.
Examples of representing median +/- interquartile 
via sio:MinimalValue and sio:MaximalValue boundary classes, and percentage association, can be viewed on our resources website.  
 \section{Applications} \label{applications}

Our study applicability applications leverage the declarative specifications of study populations in our Table 1 KG. In subsection \ref{scenarios}, we frame three scenarios of clinical relevance that mimic the decision-making of a medical practitioner to determine study applicability. Additionally, we present a cohort similarity visualization strategy in subsection \ref{viz}.
In subsection \ref{faceted browser section}, we describe a faceted browser interactive visualization tool aimed at medical practitioners. 
Moreover, as shown in Fig. \ref{cohort analytics workflow}, we include study details in our application results. Hence, we provide medical practitioners with provenance-justified results that could be used for future analyses and investigation.
\begin{figure}[hbt!]
  \includegraphics[width=0.90\linewidth, center]{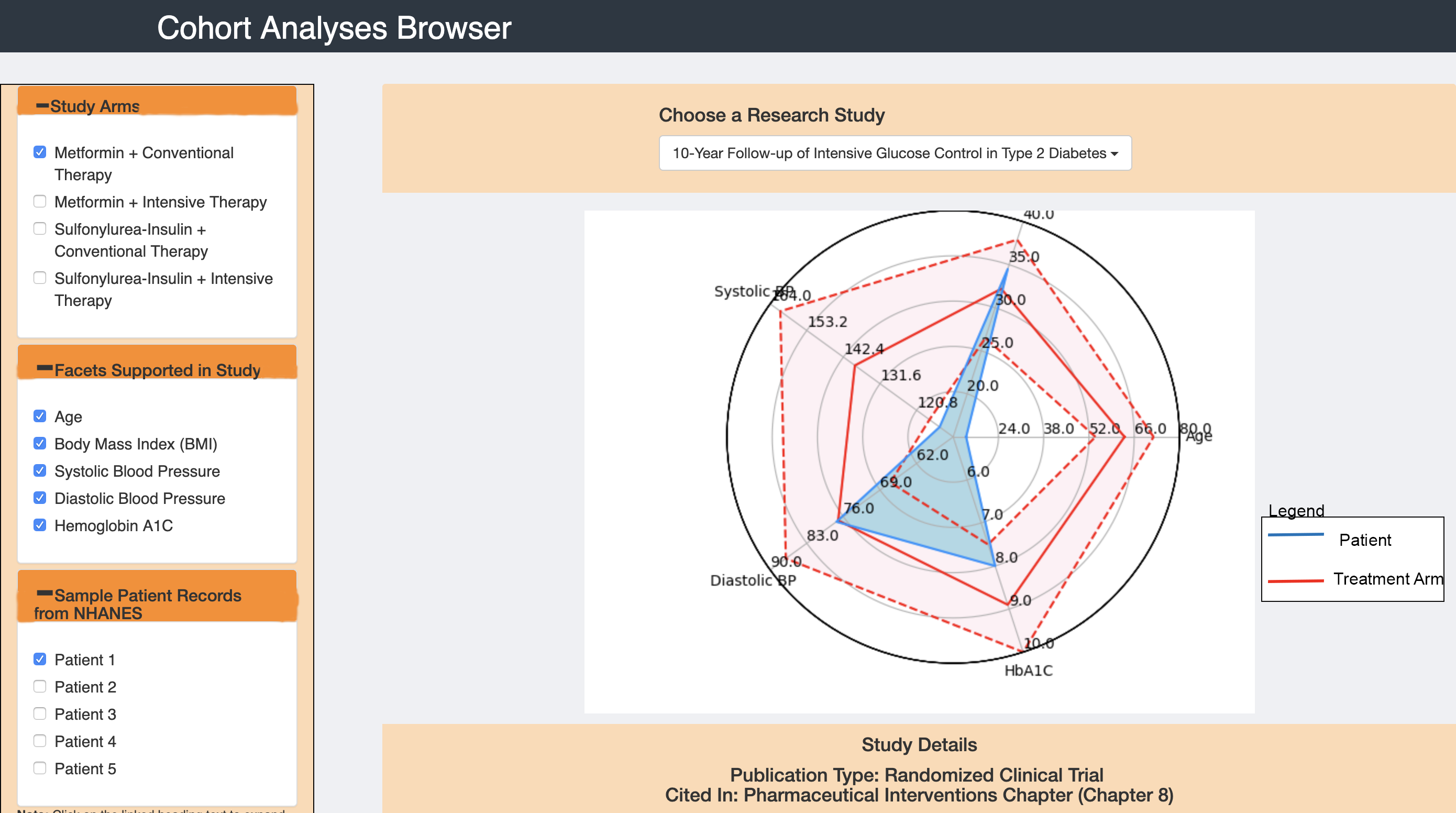}
  \caption{A snapshot of our faceted browser tool  that provides medical practitioners with the ability to customize cohort analyses. Currently, the feature facets are limited to the patient features from NHANES, that overlap with, Table 1 data. If a study doesn't contain some of these 5 features, they will be disabled.}
  \label{faceted browser}
    \end{figure}

\subsection{Population Analysis Scenarios} \label{scenarios}
As discussed
in section \ref{motivation}, there exist challenges with study biases and the varying quality in research studies. Medical practitioners need to be aware of these issues when deciding on applicable studies for their clinical population. Three scenarios of clinical relevance were suggested by our medical expert on the Health Empowerment by Analytics, Learning and Semantics (HEALS) project. Through queries to our Table 1 KG we
address a representative competency question for each of these scenarios: 
(1) Study match: Is there a study that matches this patient on a feature(s)? (2) Study limitation: Is there an absence or an underrepresentation of population groups in this study? (3) Study quality evaluation: Are there adequate population sizes and is there a heterogeneity of treatment effects among arms? Our declarative representations of Table 1s, allow us to trigger retrospective queries that combine subject characteristics (SPARQL AND clauses), various descriptive statistical representations (limited patterns of modeling as seen in section \ref{kg}),  and aggregate study arms or study cohorts (leveraging SPARQL math constructs such as SUM). Our competency questions and their SPARQL queries\footnote{\url{https://tetherless-world.github.io/study-cohort-ontology/application\#scenarioquery}} can be found on our resources website. 
\subsection{Cohort Similarity Visualizations} \label{viz}
We define cohort similarity as an analytical problem to determine the similarity or closeness of a patient to a given study population. 
We currently support determination of cohort similarity by generating visualizations, such as a star plot (Fig. \ref{faceted browser}), by overlaying features of patient records against study arm characteristics.
For the purpose of visualization, we select a few sample type-2 diabetes patient records from the National Health and Nutrition Examination Survey (NHANES)\footnote{Dataset Information Page. \url{https://wwwn.cdc.gov/nchs/nhanes/continuousnhanes/default.aspx?BeginYear=2015}}. Additionally, we adopt different visual strategies for continuous and categorical variables. In this paper, on the resources website and through our faceted browser we only support star plot visualizations for continuous variables, and we are exploring visualizations such as a pie chart for categorical variables.
Visualizations are generated on a per study arm, per patient basis, through results of SPARQL queries triggered to our Table 1 KG.
Our visualizations are built by Python plotting modules such as Seaborn\footnote{
\url{https://seaborn.pydata.org/}} and Matplotlib\footnote{
\url{https://matplotlib.org/}}, and our visualization code is made available as a resource.\\
Since our visualizations serve the purpose of being quick assessors, we design them with reduced complexity. 
Specifically, we aim for them to (1) contain  sufficient detail that is not considered overwhelming and (2) carry information such as variable ranges and the extent of the patient match, to serve as indicators for future analysis. 
\subsection{Faceted Browser} \label{faceted browser section}
We built a faceted browser tool for medical practitioners by utilizing a Python model-view-controller framework, Flask\footnote{\url{http://flask.pocoo.org/}}. On the backend (model), we utilized the RDFLib\footnote{\url{https://rdflib.readthedocs.io/en/stable}} module to trigger SPARQL queries on the ingested ontology and KG files.
Through this tool medical practitioners can interact with our Table 1 KGs, and run cohort similarity analyses on studies of their choice. They can choose from a list of studies and, subsequently, a faceted view will be rendered for the study arms of this selected study.  As seen in Fig. \ref{faceted browser}, they can also choose the variables that they would like to visualize. Hence, our prototype faceted browser interface serves as a per-study inspection tool and uses NHANES patient records to illustrate the facets. 
\section{Results}
In the Study Analysis Table \ref{population analysis summarization}, shown below,
we present a quantitative summarization of the results of each competency question (described in subsection \ref{scenarios}). Some interesting, medically relevant inferences that we output, and that are often spoken about in medical literature, include the lack of a representation of adults above 70,\footnote{
\url{https://www.statnews.com/2019/01/31/nih-rule-make-clinical-research-inclusive/}} and the lack of heterogeneity in treatment effects.\footnote{
NIH Collaboratory run grand-round presentation: \url{https://www.nihcollaboratory.org/Pages/Grand-Rounds-02-28-14.aspx}} 
We were surprised that only $6\%$ of the studies in our evaluation dataset were conducted on a large-scale, that their study arms were evenly divided, and all their study subjects were put on the basic, antidiabetes treatment of \emph{guanidines}. 
\begin{table}[tbh!]
\centering
\caption{Percentage of studies meeting the competency question criteria for the population analysis scenarios.}
\label{population analysis summarization}
\resizebox{\textwidth}{!}{
\begin{tabular}{|l|l|l|}
\hline
Question & Percentage & Population Analysis Type \\ \hline
\begin{tabular}[c]{@{}l@{}} Studies with a representation of Male African \\ American study subjects \end{tabular}
 & $75\%$  & Study Match \\ \hline
\begin{tabular}[c]{@{}l@{}} Study Arms with adults below \\ the age of 70 \end{tabular} &  $47.6\%$ & Study Limitations \\ \hline
\begin{tabular}[c]{@{}l@{}}  Studies with cohort sizes \textgreater 1000 and study \\ arm administered drugs of the guanidines family, \\ with sizes 1/3rd those of the cohort size  \end{tabular} & $6\%$  & Study Quality Evaluation  \\ \hline
\end{tabular}
}
\end{table}
We also find that the SCO ontology is epistemologically adequate for representing all Table 1s in our evaluation dataset.
We cover 360 ($\approx 17$ in each study on average) subject characteristics from 20 cited research studies,
and 28 study arm definitions. The study arm definitions included terms belonging to classes such as medical interventions, control regimes, and, less commonly occurring, diseases, dosage, year of follow-ups, and titration targets. We found that 19 cohort variables (a term we use to collectively describe interventions and subject characteristics) commonly occur across studies. 

\section{Resource Contributions}\label{sec:contributions}
We expect the following publicly available artifacts, along with the applicable documentation, to be useful resources for anyone interested in performing analysis on study populations reported in research studies. 
\begin{multicols}{2}
\begin{enumerate}    
	\item Ontologies:
  		\begin{enumerate}
    		\item Study Cohort Ontology (SCO)
  		\end{enumerate}
    \item Knowledge Graphs:
    	\begin{enumerate}
    		\item Table 1 Knowledge Graph 
  		\end{enumerate}
    \item Source Code:
    	\begin{enumerate}
    		\item MIREOT Script 
            \item Cohort Similarity Visualization
  		\end{enumerate}
    \item Data:
    	\begin{enumerate}
    		\item NHANES Patient Records
  		\end{enumerate}
    
\end{enumerate}
\end{multicols}

\section{Future Work}
Having demonstrated our ability to apply semantic techniques to make study populations visible, we plan to incorporate interdiscplinary methods to improve on a few aspects of our solution.
We have found that there exist variances in Table 1 reporting styles ranging from differences in row and column headers, table formats etc. These variances pose challenges for the scalability and automation aspects of the KG construction. 
Furthermore, often some subject characteristics and column headers require a contextual understanding for disambiguation, that is present in the unstructured body of the study. Hence, we are exploring a combination of natural language processing and semantic techniques to support an ontology-driven parsing and clean-up of Table 1 data and to develop a contextualized and medical standards compliant Table 1 KG.
Further, to ensure longevity and easy reuse of SCO, 
we plan to develop a set of tools/algorithms to predict the best-fit position for a new term in our SCO suite of ontologies. We also plan to expand and refine our set of competency questions, based on feedback from medical practitioners, and to allow for partial and fuzzy matches using query relaxation \cite{hurtado2008query} and semantically targeted analytics \cite{semantlytics}.

\section{Discussion}
We have utilized KR techniques, i.e. OWL encodings of SCO and a knowledge graph of Table 1 content to model and expose descriptions of study populations in an attempt to make scientific data more accessible. Further, we have utilized our semantic modeling to support analytical use cases to determine study applicability.Our evaluation dataset currently is solely comprised of type-2 diabetes research studies.  We have kept our descriptions and examples minimally domain specific. We believe that our ontology and KG documentation can serve as resources for researchers interested in the pan-disease analysis of study populations.\\
Our ontology, SCO, is developed using best-practice ontology principles, some of which are listed at \cite{widocobestpractices}. Specifically, we
reuse SIO properties and do not define any 
new properties. We reuse classes from a limited yet standard set of biomedical ontologies in order to increase the interoperability of SCO. \\
There have been attempts at improving the reporting of Table 1s in the medical community, such as the Table 1 project \cite{duketable1project}. However, they have been confined to the identification of desirable properties for standardization. Our semantic solution presented in this paper, that at its heart utilizes a KR approach, is a step towards achieving this standardization.
This can be seen in Listing 1.1 where we have presented an RDF snippet representing fundamental building blocks of our Table 1 KG, i.e. our modeling of collections, subject characteristics, and statistical measures. These identified patterns are reused as templates to realize the association of various variables with study populations reported in Table 1s. \\
Our Table 1 KGs allows us to address study applicability scenarios motivated from medical literature and to support visualizations that clearly depict cohort similarity. By these capabilities, we demonstrate how our solution addresses our use case of determining study applicability. 
We believe there is potential for this work to be reused by researchers performing study population analyses. Also in this paper we make assumptions on the content a medical practitioner might want to see, and, from a medical practitioner user survey we are conducting, we will incorporate feedback on their additional requirements.\\
Our solution does not address or include support for the modeling of study eligibility criteria, i.e. inclusion and exclusion criteria. 
However, we reuse metadata expression terms from Dublin Core Terms (DCT) to include a link to registries such as ClinicalTrials.gov or International Standard Randomised Controlled Trial Number (ISRCTN),\footnote{\url{http://www.isrctn.com/page/about}}
where the criteria is made available as a part of the study data. We 
 expect that the SCO vocabulary is sufficient to express the criteria, but since we are still investigating the merge of the criteria with the Table 1 content, we defer it to future work. \\
Finally, all the resources that we listed in section \ref{sec:contributions}, are made publicly available in a Github repository and the ontology is hosted on Bioportal. SCO is released under the Apache 2.0 license specification. Our resources will be maintained periodically by the authors.
\section{Conclusion}
We have presented a prototype KR system that can be used to model study populations, to aid in the assessment of study applicability. Our model is tailored around 
use cases aimed at assisting medical practitioners in the treatment of complex patients and who also often require 
``efficient-literature searching'' \cite{masic2008evidence} capabilities. We presented a solution to make descriptions of study populations more accessible for quick decision-making. We believe that the resources we release, especially SCO, can serve as an extensible schema to represent population descriptions across diseases. We have demonstrated the adequacy of the ontology through a set of what we believe are representative applications supporting a range of use cases contributed by our medical expert.  We plan to continue our outreach and ontology reuse in additional diverse evidence-based medicine application settings. 

\section*{Acknowledgements}
This work is partially supported by IBM Research AI through the AI Horizons Network. We thank our colleagues from IBM Research, Dan Gruen, Morgan Foreman and Ching-Hua Chen, and from RPI, John Erickson, Alexander New, and Rebecca Cowan, who greatly assisted the research.

\bibliographystyle{splncs04}
\bibliography{references}
\end{document}